\documentclass[twocolumn]{aa}
\usepackage{graphicx}
\usepackage{lscape}
\def\zl{z_{\rm lens}}
\def\kms{\rm km\ s^{-1}}
\def\kmsmpc{\rm km\ s^{-1}\ Mpc^{-1}}
\begin{document}
\newcommand{\obj}{PKS~1830$-$211}
\title{Confirmation of two extended objects along the line of sight to
  PKS~1830-211 with ESO-VLT adaptive optics imaging\thanks{Based on
    observations obtained at the European Southern Observatory using
    the Very Large Telescope, Cerro Paranal, Chile (ESO Program
    ID 71.A-0401(A), PI: G.  Meylan).  Based in part on data obtained
    with the NASA/ESA Hubble Space Telescope (Program \#9744, PI: C.~S.
    Kochanek) and extracted from the data archives at the Space
    Telescope Science Institute, which is operated by the Association
    of Universities for Research in Astronomy, Inc., under NASA
    contract NAS~5-26555.}}

\author{G. Meylan \inst{1}
\and F. Courbin   \inst{1}
\and C. Lidman    \inst{2}
\and J.-P. Kneib \inst{3}
\and L.E. Tacconi-Garman \inst{4}}

\institute{
Laboratoire d'Astrophysique, Ecole Polytechnique F\'ed\'erale de
Lausanne (EPFL), Observatoire, CH-1290 Sauverny, Switzerland
\and
European Southern Observatory, Casilla 19001, Santiago, Chile
\and
LAM, Observatoire Astronomique de Marseille-Provence, Traverse du
Siphon - B.P.8, 13376 Marseille Cedex 12, France
\and
European Southern Observatory, 
Karl-Schwarzschild-Stra$\ss$e 2, D-85748 Garching bei M\"unchen, Germany
}

\date{Received / Accepted }

\abstract{We report on new high-resolution near-infrared images of the
  gravitationally  lensed radio  source  \obj, a quasar  at $z=2.507$. 
  These  adaptive   optics observations, taken   with  the  Very Large
  Telescope (VLT), are  further improved through image  deconvolution. 
  They confirm the presence of a second object along the line of sight
  to the quasar, in addition  to the  previously known spiral  galaxy. 
  This additional object is  clearly extended in our images.  However,
  its  faint   luminosity does  not  allow   to infer  any photometric
  redshift.  If  this galaxy is located  in the foreground of \obj, it
  complicates the  modeling of this  system and decreases the interest
  in using  PKS~1830--211 as a means to  determine $H_0$  via the time
  delay between the two lensed images of the quasar.
  
  } \titlerunning{Multiple galaxies along the line of
  sight to PKS~1830-211} \maketitle

%%%%%%%%%%%%%%%%%%%%%%%%%%%%%%%%%%%%%%%%%%%%%%%%%%%%%%%%%%%%%%%%%%%%%%%%
%%%%%%%%%%%%%%%%%%%%%%%%%%%%%%%%%%%%%%%%%%%%%%%%%%%%%%%%%%%%%%%%%%%%%%%%

\section{Introduction}

\obj\ is one of the first quasars that was found to  be lensed.  It is
seen in the radio as a full Einstein ring connecting two quasar images
(Rao \& Subrahmanyan  1988, Subrahmanyan et~al.  1990, Jauncey  et~al. 
1991).  The time delay between the quasar  components is known from an
18-month radio monitoring with ATCA  at 8.6~GHz, providing $\Delta t =
26 \pm 5$ days  (Lovell et~al.  1998).   While the quasar is bright in
the  radio, it  is   much fainter in   the  optical.  It  is  only  in
near-infrared that both quasar images  can be identified, either using
ESO/KECK $I, J, K$ images (Courbin  et~al.  1998), or the Hubble Space
Telescope (HST) with the NICMOS instrument (Leh\'ar et~al.  2000).

%  Even with HST  data, these observations were challenging,  due
%to the fact  that \obj\ is located  very close to the galactic  plane,
%hence in a field crowded with foreground stars.

The redshift of the  background quasar is  known from IR spectroscopy,
$z=2.507$, using the redshifted H$\alpha$ emission line (Lidman et al.
1999),  while the identification    of  the lensing galaxy  has   long
remained problematic.  Indirect identification of the lens was already
possible  in early millimeter observations  of \obj,  thanks to strong
absorption lines  at  $z=0.89$ that were found  in  the spectra of the
quasar images (Wiklind \& Combes 1996).   The direct identification of
a lensing object, in the form of a spiral galaxy, is much more recent:
using HST optical and IR images Courbin et  al.  (2002; hereafter C02)
and  Winn et  al.  (2002;  hereafter   W02) identify a  spiral  galaxy
between the  quasar images.  The  $V-I$ color of this galaxy indicates
that it is probably associated with the $z=0.89$ millimeter absorber.

In addition to the spiral  lens found by both  C02  and W02, C02  also
noticed the  presence  of a  second  diffuse light source between  the
quasars,  which is already   pointed out by  Leh\'ar  et al.   (2000). 
While Leh\'ar et al.  (2000) and C02 identify this object as a lensing
galaxy, W02  assume  that  it is   an  image  artifact  or  a  feature
associated  with either the quasar  or the more obvious lensing spiral
galaxy.  This second  object, ``Lens G'', is faint  and very close  to
the center  of  the spiral  galaxy and  is  seen only in  the  IR (see
Fig.~1). Only deep near-infrared  imaging with Adaptive Optics (AO) on
a 10-m  class telescope can  significantly improve the  quality of the
current observations.  This paper presents  such new near-infrared  AO
images,  obtained  with  the  VLT.   The  data, further improved  with
deconvolution techniques, allow us to confirm unambiguously the second
object,   ``Lens G'', located between   the two  quasar images, without,
however, being able to put any constraints on its redshift.

%%%%%%%%%%%%%%%%%%%%%%%%%%%%%%%%%%%%%%%%%%%%%%%%%%%%%%%%%%%%%%%%%%%%%%%%
%%%%%%%%%%%%%%%%%%%%%%%%%%%%%%%%%%%%%%%%%%%%%%%%%%%%%%%%%%%%%%%%%%%%%%%%

\begin{figure}[t!]                                            
\centering
\includegraphics[width=8.5cm]{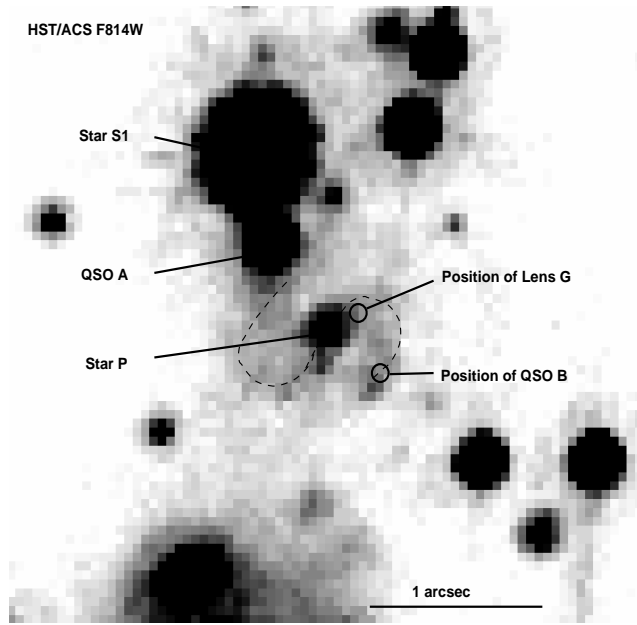}          
\caption{Part of an HST/ACS F814W image obtained by the CASTLE group.
  This  deep 10,390  s  exposure shows the   spiral lens (dashed line)
  pointed out by C02 and W02. Component A of the quasar is conspicuous
  while the component  B is  obscured by  one  the spiral arms of  the
  lensing galaxy, which center is   not unambiguously identified:  W02
  consider  Star P as the  bright  unresolved nucleus  of the  spiral,
  while  C02 consider  it as  a  mere star.   The position  of Lens~G,
  confirmed   by the present  VLT  AO observations  is indicated.  All
  objects are labeled as in C02.  North is up and East to the left.}
\label{HST_ACS} 
\end{figure}

\section{VLT adaptive optics observations and reduction}

Deep $J$, $H$ and $K_{\mathrm  s}$-band images of \obj\ were obtained
at  the ESO Paranal  Observatory with the  near-infrared camera CONICA
(the  COude Near-Infrared CAmera), which is  mounted on  the AO system
NAOS (Nasmyth  Adaptive  Optics  System).   Both CONICA and   NAOS are
mounted on the Nasmyth focus of VLT  UT4 (Yepun). The $J$ and $H$-band
data were taken during  the night of 27 May,  2003 and the $K_{\mathrm
  s}$-band  data were taken during the   night of 3rd September, 2003. 
The   pixel  size of  the  instrument,  0\farcs02704$\pm$0\farcs00007,
allows the acquisition  of well sampled  data at the diffraction limit
of the telescope.

Until the   laser guide star  becomes available  at  the VLT, a nearby
bright star is required to  carry out the  wavefront correction by the
AO system.  Thanks to  the low galactic latitude  of \obj,  a $V=13.5$
star is  available 7\arcsec\ to  the North-East of  the quasar.  It is
the same  star that was used  for the  $K$-band Gemini AO observations
presented in   Fig.~1 of C02.   The seeing  during  these observations
results in a final AO-corrected resolution that  is always better than
0\farcs1.

Standard reduction techniques are applied  to the data.  They  include
dark  subtraction,  flat fielding   using   twilight  flats,  and  sky
subtraction.  The sky  frames are computed  using the  14 images taken
nearest in time to the frame  considered.  The reduced frames are then
registered and combined, rejecting the  highest and the lowest  pixels
in  the stack. Cosmic rays and  pixels  with variable dark current are
removed, since dithering is  applied   during the observations.    The
details of the observations are summarized in Table~\ref{obslog}.

\begin{table}[t!]
\caption[]{Summary of the VLT adaptive optics observations of
  \obj. The length of the individual exposures is DIT seconds.
  NDIT exposures are averaged for each of the NDITHER telescope
  positions. The value of the AO-corrected resolution is the 
  FWHM of the core of the adaptive optics PSF.}
\label{obslog}
\begin{flushleft}
\begin{tabular}{lcccr}
\hline\hline 
Filter &  Airmass   & NDIT$\times$DIT        & Resolution  &  Exp. Time \\
       &  (mean)    & $\times$NDITHER         & (FWHM)   &   (total)\\
\hline
J      & 1.10       & 3$\times$40$\times$37  &  0\farcs085  &    4440 s \\
H      & 1.20       & 3$\times$20$\times$104  &  0\farcs070  &   6240 s \\
Ks     & 1.03       & 2$\times$30$\times$14  &  0\farcs097  &     840 s \\
\hline 
\end{tabular}
\end{flushleft}
\end{table}

%%%%%%%%%%%%%%%%%%%%%%%%%%%%%%%%%%%%%%%%%%%%%%%%%%%%%%%%%%%%%%%%%%%%%%%%
%%%%%%%%%%%%%%%%%%%%%%%%%%%%%%%%%%%%%%%%%%%%%%%%%%%%%%%%%%%%%%%%%%%%%%%%

\section{Deconvolution and astrometry}

\begin{figure*}[t!]                                            
\centering
\includegraphics[width=14.0cm]{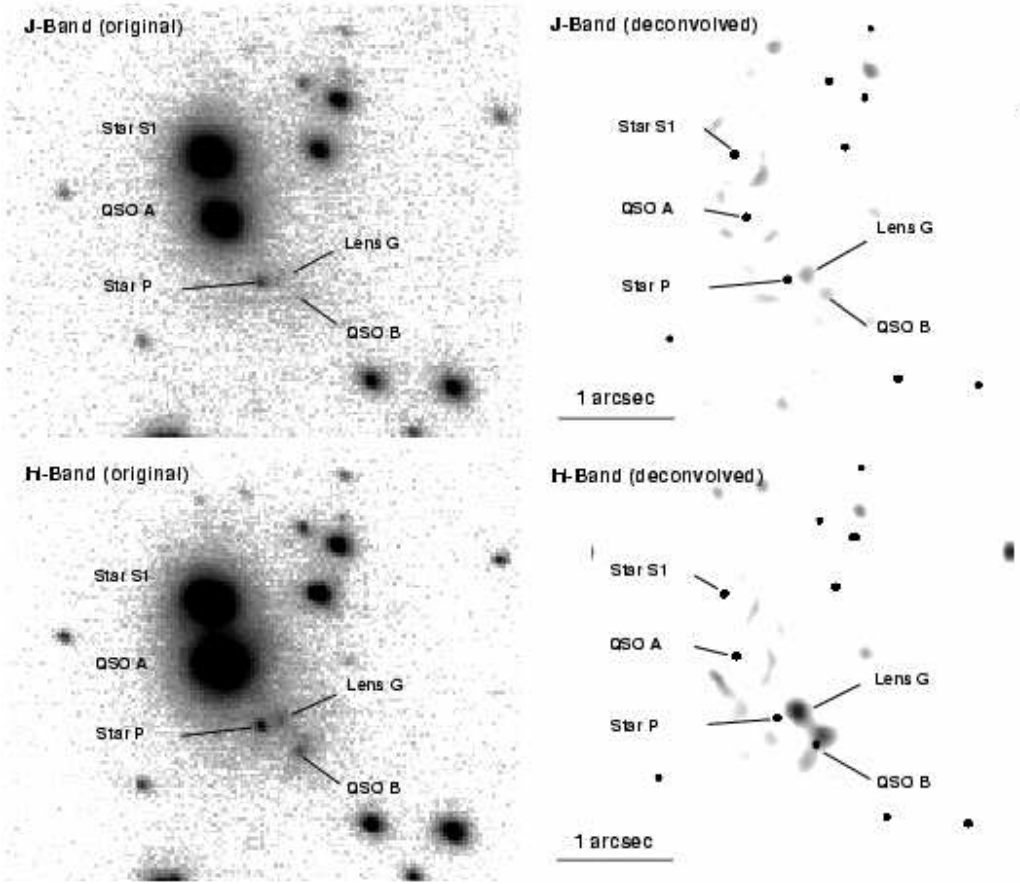}          
\caption{The combined VLT NACO images of \obj\ are shown on the left
  side of the figure, where all objects are labelled as in C02. Lens G
  is at the  detection limit in  the $J$-band, but is clearly detected
  in $H$.  The  deconvolved  images are shown  on  the  right,  with a
  resolution of 0\farcs027.  In addition to Lens G, an extended object
  near   the  Image  B of    the  quasar   is  seen   in the  $H$-band
  deconvolution, and may  correspond to the  lensed host galaxy of the
  quasar source.  Both in the reduced and deconvolved images, the grey
  scale  displays  all    intensity   levels  above   2-$\sigma_{\rm sky}$. 
  Artifacts due  to the PSF  spatial  variations across the CONICA  camera,
  within a    radius of 0.5\arcsec\  around   bright point sources are
  significant, but do not affect Lens G.}
\label{VLT_NACO} 
\end{figure*}

Two objects are striking in the new AO data (Fig.~\ref{VLT_NACO}, left
panels), especially in the $H$ filter: Star P from C02, and Lens G, at
the position  already pointed out by C02  and Leh\'ar  et al.  (2000). 
The  spiral lens seen  in the HST  optical images  is too  faint to be
detected in these near-infrared images. 

While Lens~G    is already nicely   confirmed  by the  adaptive optics
images, image deconvolution  allows  one to  go further and  to derive
extremely precise astrometry.  The MCS deconvolution algorithm (Magain
et al.   1998)  is used, reaching  a  resolution of 0\farcs027 in  all
three  near-infrared  filters.    The data   are    decomposed into  a
point-source (stellar and  quasar images)  and extended ``channels''.  
The lensing galaxies    and the quasar host  are   represented in this
latter channel.   We note  an extended object   in the $H$-band  image
surrounding the quasar Image  B. It can be either  the lensed  host of
the quasar, or one more  intervening object along  the line of sight.  
The pixel scale in the deconvolved image has been improved by a factor
of  two  over  the   original  data, i.e.   the  new   pixel  size  is
0\farcs01352.  The deconvolved $J$  and $H$-band images are displayed
in Fig.~\ref{VLT_NACO} (right panels).  Lens~G  is also visible in the
shallower $K_{\mathrm s}$-band data.

The quality of the deconvolved images is mainly  limited by the strong
PSF variation  across the field  of  view.  As a result,  and although
several stars are available in order to carry out the PSF calculation within
a circle of 10\arcsec\ around \obj, small artifacts  are seen close to
the  quasar images at faint  levels.  This  effect is significant only
near the bright objects S1 and quasar Image A  and does not affect the
deconvolution at the location of the much fainter Lens~G.  These faint
artifacts,  shown in  log  scale on   Fig.~\ref{VLT_NACO}, are  a  few
percent  of the peak  intensity of Lens~G.  The PSF wings are modelled
over  an area which  has the   same size  as  the  field used for  the
deconvolution.

\begin{figure}[h!]                                            
\centering
\includegraphics[width=8.5cm]{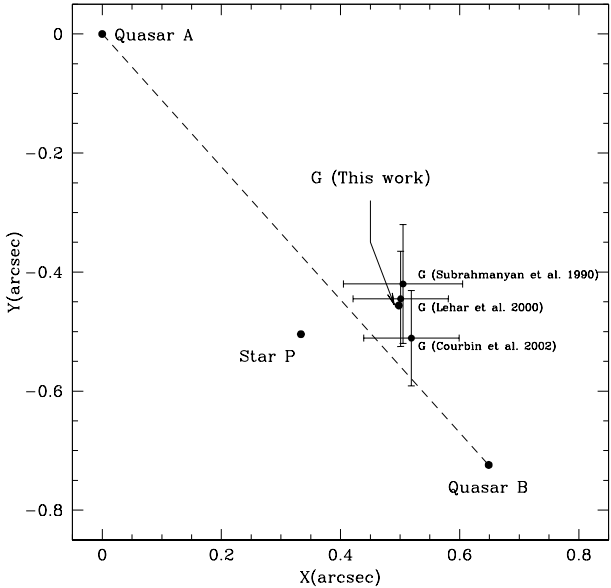}          
\caption{Summary of the astrometry of \obj, relative to the quasar
  Image A. The position of Lens~G is compared with the position found
in C02 and with that of ``Object E'' seen in the radio by Subrahmanyan
et al. (1990).}
\label{astrometry} 
\end{figure}

We summarize in Table~\ref{astrometry_table} and Fig.~\ref{astrometry}
the  astrometry of the different objects.   All objects are visible in
the $J$,  $H$  and $K_{\mathrm  s}$-bands.    The quasar  Image B   is
nevertheless very faint in   $J$, so we  do  not  use this filter   to
constrain the position of component B.   The astrometry presented here
is the mean of the astrometry derived either in $H+K$ (quasar B) or in
$J+H+K$ when the  objects are visible in  all three bands.  The quoted
error bars are the errors on this mean  value, i.e. the square root of
the variance of the  different measurements divided  by $N$, where $N$
is the number of filters where a measurement is possible.

Flux calibration is carried out  using several standard stars,  except
for  the  $K_{\mathrm  s}$-band data.  The   photometry  for all point
sources is   done by  integrating  the flux   in  the {\it  analytical
  gaussian}    profiles  used  to    model    point   sources  in  the
MCS-deconvolved images.  Extended objects such  as Lens~G are measured
through 0\farcs2 diameter apertures on  the quasar-subtracted images.  
All photometric measurements are given in Table~3.

We attempt to model Lens~G as a point source when carrying out the MCS
deconvolution.  The result  shows significant  residuals in the   {\it
  extended channel}    of  the image, demonstrating    that  Lens~G is
resolved in the AO data.  Only  the high signal-to-noise $H$-band data
allow to carry out  this  experiment.  Star~P remains  unresolved (see
Fig.~\ref{VLT_NACO}), following this test.

%%%%%%%%%%%%%%%%%%%%%%%%%%%%%%%%%%%%%%%%%%%%%%%%%%%%%%%%%%%%%%%%%%%%%%%%
%%%%%%%%%%%%%%%%%%%%%%%%%%%%%%%%%%%%%%%%%%%%%%%%%%%%%%%%%%%%%%%%%%%%%%%%

%%%%%%%%%%%%%%%%%%%%%%%%%%%%%%%%%%%%%%%%%%%%%%%%%%%%%%%%%%%%%%%%%%%%%%%%
%%%%%%%%%%%%%%%%%%%%%%%%%%%%%%%%%%%%%%%%%%%%%%%%%%%%%%%%%%%%%%%%%%%%%%%%

\section{Discussion and conclusions}

Our deep high-resolution VLT adaptive optics observations clearly
confirm the existence of the second extended object ``Lens~G''. We
also confirm that ``Star P'' is unresolved at the resolution of the
present AO images.

Without any redshift measurement, it is  hard to evaluate the relative
importance of  the two extended   objects within the Einstein ring  of
\obj.  The only clear  clue about a  lens redshift in \obj\ comes from
the molecular absorption   lines seen in  the  radio spectrum  of  the
quasar images,  suggesting $\zl=0.89$ (Wiklind  \& Combes 1996).  This
redshift is more likely  associated with the  spiral lens rather  than
Lens~G.   First, the HST images  (Fig.~1)  show that  the arms of  the
spiral  galaxy are  passing  right  in front  of   the quasar  images,
especially B. Second,  the molecular absorption lines  seen in the two
quasar images display a velocity difference  of $\Delta V=147\,\kms $,
typical  of a rotating  disk with a small  inclination with respect to
the plane of the sky.   Finally, the $V-I$  color of the spiral  (W02,
C02) is indicative a spiral galaxy  that has a  redshift that is close
to $z=1$.

The redshift of  Lens~G is much  less certain, as photometric data for
this object are available only in $J$ and $H$,  and with large errors. 
Lens~G is  located only 0\farcs2  away  from the  center of the spiral
disk. Since this is only 1 kpc at $z=0.89$, Lens~G could  be part of the
spiral galaxy seen in the HST images, if both objects are at $z=0.89$.

If Lens~G is not simply part  of the spiral galaxy,  but is rather the
optical counterpart of ``Object E'' seen in  the radio by Subrahmanyan
et al.  (1990),  then the $K$ vs.   redshift (or approximately $H$ vs. 
redshift) relation applies and indicates $\zl=1.1-1.2$.  This redshift
is hard to reconcile with  lens models.  If we take  the models by W02
using   a single lens  centered at  the position of  Lens~G, we obtain
values for H$_0$ of about 80\,$\kmsmpc$.  Taking into account the fact
that W02 compute their models  for  $\zl=0.89$ and not  $\zl=1.1-1.2$,
the  redshift-corrected   values   for    H$_0$ rise  above
100\,$\kmsmpc$.    Acceptable   values   for   H$_0$,   in  the  range
$50<H_0<80$\,$\kmsmpc$ are found when the redshift  of Lens~G is about
$\zl=0.4-0.5$.

%Following the idea  that two lenses must  be considered to model \obj,
%we have attempted to model the object  using the non parametric models
%of   Saha   \&   Williams    (2004).    Using   a    known  cosmology,
%($\Omega_{\Lambda}$=0.7,   $\Omega_{m}$=0.3, H$_0$=65 \,$\kmsmpc$) and
%the measured  time delay for \obj, it  is possible to construct a lens
%model composed of  a symetric part   plus an assymetric residual.   We
%computed  two models, one using a  lens centered on Lens~G and another
%one using a lens centered on Star P.  The lens redshift was
%fixed at  $\zl$=0.89. In both  cases, the  residual  part of the  lens
%model representing deviation  from central symetry  is negligible, and
%is predicted at random places between the quasar images, as one varies
%the  exponent of the power law  profile in the symetrical component of
%the model.

\begin{table}[t!]
\caption[]{Astrometry of \obj.
All positions are given relative to the quasar Image A along with their
1-$\sigma$ error bars. The measurements have been
transformed to match the coordinate system of NICMOS (Leh\'ar et
al. 2000), by using 8 stars common to the AO and NICMOS
fields of view.}
\label{astrometry_table}
\begin{flushleft}
\begin{tabular}{lcc}
\hline\hline 
Object    &      X                 &  Y   \\
          & (arcsec)               & (arcsec) \\
\hline
Quasar A  &  0.0                 & 0.0  \\
Quasar B  & $+$0.649$\pm$0.001   & $-$0.724$\pm$0.001 \\
Star P    & $+$0.333$\pm$0.002   & $-$0.504$\pm$0.002 \\
Lens G    & $+$0.498$\pm$0.004   & $-$0.456$\pm$0.004 \\
\hline 
\end{tabular}
\end{flushleft}
\end{table}

%\begin{table}[t!]
%\caption[]{Photometry of the objects in the vicinity of \obj.}
%\label{photometry_table}
%\begin{flushleft}
%\begin{tabular}{lccc}
%\hline 
%Object    &       J         &      H          &    K   \\
%\hline
%\hline
%Quasar A  & 17.98$\pm$0.05  &  16.31$\pm$0.05 &  14.58$\pm$0.05 \\
%Quasar B  & $>$23.0         &  22.07$\pm$0.10 &  17.77$\pm$0.10 \\
%Star S1   & 17.51$\pm$0.05  &  16.67$\pm$0.05 &  16.30$\pm$0.10 \\
%Star P    & 21.70$\pm$0.10  &  20.88$\pm$0.05 &  20.05$\pm$0.20 \\
%Lens G    & 22.70$\pm$0.20  &  21.20$\pm$0.10 &  20.50$\pm$0.20 \\
%\hline 
%\end{tabular}
%\end{flushleft}
%\end{table}

\begin{table}[t!]
\caption[]{Photometry of the objects in the vicinity of \obj, along
  with their 1-$\sigma$ error bars. An additional error bar of 0.05
  magnitude on the photometric zero point is not included here.}
\label{photometry_table}
\begin{flushleft}
\begin{tabular}{lcc}
\hline\hline 
Object    &       J         &      H         \\
\hline
Quasar A  & 18.04$\pm$0.05  &  16.36$\pm$0.05  \\
Quasar B  & $>$23.0         &  22.12$\pm$0.10  \\
Star S1   & 17.57$\pm$0.05  &  16.72$\pm$0.05  \\
Star P    & 21.76$\pm$0.10  &  20.93$\pm$0.05  \\
Lens G    & 22.76$\pm$0.20  &  21.25$\pm$0.10  \\

\hline 
\end{tabular}
\end{flushleft}
\end{table}

Finally, we can not exclude the  possibility that Lens~G is behind the
$z=2.507$ quasar and is not acting as a lens at all.  Further study of
\obj\  will require spectroscopy of both  Lens~G and  the spiral lens,
which,  given the  required  spatial  resolution  and depth, is   only
possible from space.  Spectroscopy of Star P, which could be the bulge
of the spiral lens  (W02,  C02), remains  within  the reach of   large
ground-based telescopes with AO.

\begin{acknowledgements}   
  The authors  would  like to thank Dr.    Prasenjit Saha  for helpful
  discussions and the referee, Dr. Joshua Winn,  for his very positive
  and  constructive remarks. GM  and FC  acknowledge  support from the
  Swiss National Science Foundation (SNSF).
\end{acknowledgements}

%%%%%%%%%%%%%%%%%%%%%%%%%%%%%%%%%%%%%%%%%%%%%%%%%%%%%%%%%%%%%%%%%%%%%%%%
%%%%%%%%%%%%%%%%%%%%%%%%%%%%%%%%%%%%%%%%%%%%%%%%%%%%%%%%%%%%%%%%%%%%%%%%

\end{document}